\documentclass{article}
\usepackage{graphicx} 

\usepackage[T1]{fontenc}
\usepackage{microtype}
\usepackage{booktabs}
\usepackage{array}
\usepackage{tabularx}
\usepackage{enumitem}
\usepackage{xcolor}
\usepackage{tikz}
\usetikzlibrary{positioning,fit,arrows.meta,calc,shapes.geometric}
\usepackage{hyperref}
\usepackage{xurl}

\usepackage[accepted]{mlsys2025}
\makeatletter
\renewcommand{\@pa}[1]{}
\gdef\printAffiliationsAndNotice#1{{\let\thefootnote\relax\footnotetext{\hspace*{-\footnotesep}%
  Amazon. Correspondence to: \mlsyscorrespondingauthor@text.

  \ \\
  \Notice@String}}}
\renewcommand{\Notice@String}{Preprint.}
\makeatother

\mlsystitlerunning{Plumbline: When the Loss Lies, Validating LLM Training Correctly}

\usepackage{wasysym}
\usepackage{makecell}
\usepackage{longtable}
\usepackage{placeins}

\newcommand{\appBcite}{}
\newcommand{\appAconv}{the run's records show both}
\newcommand{\appAsent}{}

\makeatletter
\newcounter{appprobno}
\expandafter\def\csname app@skip@not reported\endcsname{}
\newcommand{\appprob}[8]{\ifcsname app@skip@#6\endcsname\expandafter\@gobble\else\expandafter\@firstofone\fi
  {\stepcounter{appprobno}\theappprobno & #2 & #3 & #4 & #5 & #6 \\*
   & \multicolumn{5}{>{\raggedright\arraybackslash\hyphenpenalty=10000\exhyphenpenalty=10000}p{\appprobw}@{}}{\textit{#8}} \\ \addlinespace[6pt]}}
\makeatother

\newcommand{\appbegintab}{\begin{tabular}{@{}rlllll@{}}}
\newcommand{\appprobw}{0.925\textwidth}

\AtBeginDocument{\hypersetup{pdfsubject={},pdfkeywords={Machine Learning, training correctness, validation, LLM training, agents}}}
\begin{document}

\twocolumn[
\mlsystitle{Plumbline: When the Loss Lies, Validating LLM Training Correctly}

{\begin{mlsysauthorlist}
\mlsysauthor{Ruochen Wen}{amzn}
\mlsysauthor{Hongyang Wang}{amzn}
\mlsysauthor{Siwei Li}{amzn}
\mlsysauthor{Wenbo Chen}{amzn}
\mlsysauthor{Xian Li}{amzn}
\mlsysauthor{Jingguo Liu}{amzn}
\mlsysauthor{Leo Fan}{amzn}
\mlsysauthor{Yiming Yan}{amzn}
\mlsysauthor{Qian Qu}{amzn}
\mlsysauthor{Zheshen Wang}{amzn}
\mlsysauthor{Benoit Dumoulin}{amzn}
\end{mlsysauthorlist}}

\mlsyscorrespondingauthor{Ruochen Wen\newline}{\mbox{ruochw@amazon.com}, \mbox{ruochenwen@gmail.com}}
\mlsyskeywords{Machine Learning, training correctness, validation, LLM training, agents}

\vskip 0.2in

\begin{abstract}
Correctness is a prerequisite for training large language models (LLMs) at scale, yet it is hard to ensure: a fault can sit in any part of a run, and how it shows differs across frameworks and the libraries they depend on. Some failures are loud and some are silent, and both take time to detect and trace. A single fault can waste a long run: the run crashes late, or the model scores lower only at the final evaluation. We present Plumbline, a correctness validation system for LLM training, used in production at Amazon. Plumbline checks each part of the training lifecycle, from the data a run trains on to the model it serves, down to the new layers each model brings and up to whole runs. Its diagnosis reuses these checks to trace failures to their root causes. Both rest on a systematic correctness methodology designed to carry across models and frameworks. Plumbline supports two agentic use cases: Day-0 onboarding, which validates the training of a newly released model before teams rely on it, and root-cause tracing of training failures. Across 3 frameworks and 5 models, it found 61 problems, 32 of them silent. Of these, 42 are fixed or reported upstream, 8 of them first raised by us. We further group the 61 into seven failure patterns that show where training breaks, and report how problems recur across frameworks and models.
\end{abstract}
]

\printAffiliationsAndNotice{}

\providecommand{\OWNER}[1]{\textcolor{violet}{[\textbf{Owner to pick:} #1]}}
\providecommand{\RECOUNT}{}
\section{Introduction}
\label{sec:intro}

Teams that train large language models (LLMs) commonly build on released model checkpoints and open-source training
frameworks. Errors in such training can arise in the user's code and configuration, in the framework, and in the
libraries and hardware beneath it~\cite{hu2024acme,l4}. These errors are costly to resolve: on one production
platform, diagnosing an escalated training failure took 34.7 hours on average and more than a day in 41.9\% of
cases~\cite{l4}.

A failure seldom points to its root cause. A fault can surface far from where it lies, or stay silent: the loss still
falls while the run trains or serves the wrong model. Whether a fault shows depends on how a run combines its model,
settings, framework and libraries (\S\ref{sec:motivation}). Tracing a fault can take reruns at the same scale.

\begin{figure*}[t]
\centering
\includegraphics{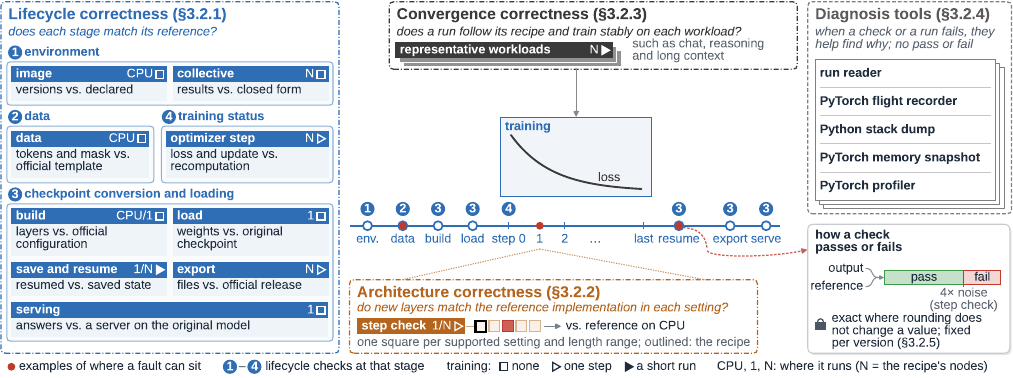}
\caption{Plumbline's checks along one training run, grouped by where a fault sits, and its diagnosis tools. Each check is labeled with what it compares, the training it needs and where it runs. The bottom row explains the markers.}
\label{fig:system}
\end{figure*}

These costs argue for validating training correctness before a long run, instead of discovering errors after a run
fails or a trained model underperforms. A fault can sit in any part of a run and need not show in the loss, so we
check each part on its own against what correct training produces. A failed check then also shows where to look,
and tracing can reuse the checks at small scale instead of rerunning the job. We built Plumbline on this
idea (Fig.~\ref{fig:system}). Our contributions are:
\begin{itemize}
\item Plumbline, a correctness validation system for LLM training. Its checks cover a run at three levels: each stage of its lifecycle, from its data to the model it serves; one training step under each supported setting; and whole runs. Its diagnosis reuses the checks at smaller scale, with tools that record what a run did, to trace a failure to its root cause.
\item A systematic correctness methodology designed to carry across models and frameworks. It follows two principles. Under correctness first, each check tests the training a user would run against a reference not produced by the code under test, under rules fixed per system version, with bounds above measured noise. Under modularity, each check tests one part on its own, so the checks run alone or together as a full scan, and a new model or framework changes how a check connects to it, not what it tests. The same checks and rules ran on all 13 framework--model combinations.
\item Plumbline is used in production at Amazon in two agentic use cases: Day-0 onboarding of newly released models and root-cause tracing of training failures. Across 3 frameworks and 5 models, Day-0 onboarding found 61 problems, 32 of them silent, and tracing located the cause of each. Of these, 42 are fixed or reported upstream, 8~of them first raised by us. We group the problems into seven failure patterns that show where training breaks, and report how problems recur across frameworks and models.
\end{itemize}
\section{Motivation}
\label{sec:motivation}
\providecommand{\RECOUNT}{}

Open-source training frameworks such as Megatron-LM, DeepSpeed and TorchTitan~\cite{shoeybi2019megatron,rasley2020deepspeed,liang2025torchtitan} run LLM training as a distributed job: they split a model and its data across many GPUs with several forms of parallelism and train it in mixed precision. The frameworks and the libraries beneath them are developed by different teams and released at different times~\cite{huang2023dependency}. New model architectures keep adding layer types for each framework to support~\cite{deepseekv3,minimax2025m1}. A training job therefore combines code that different teams change on their own schedules.

In one study of dependency bugs in deep-learning stacks, half were exposed in a different dependency from the one that introduced them~\cite{huang2023dependency}. On a Microsoft platform, 35\% of 400 sampled program failures came from differences between a developer's environment and the platform's~\cite{zhang2020icse}. Whether a fault appears therefore depends on how a run combines its parts, so a part that passes its own tests can still fail in a particular run.

In a study of 1,000 fixed bugs in four deep-learning frameworks, 24.3\% produced incorrect results without a crash~\cite{chen2023frameworkbugs}. A fault of this kind shows up late, when the trained model is evaluated or served. On the Microsoft platform, failures during model evaluation were 24\% of 400 sampled program failures but used 81\% of their GPU time~\cite{zhang2020icse}.

In a study of 428 LLM training failures on another production platform, 89.9\% needed manual log analysis~\cite{l4}. Many crashes appear as generic communication errors~\cite{dong2025c4,kokolis2025reliability}, and when a failure does not reproduce locally, developers resubmit modified jobs by trial and error~\cite{zhang2020icse}. Tracing a failure to its cause is thus slow and largely manual.

Model libraries and training frameworks implement each new model, and their tests check that implementation~\cite{mbridge-newmodel,msswift2026mcustom,hf-testing,torchtitan-debugging}. Users then train the model with their own settings, data and checkpoints, so whether a particular run is correct depends on these choices as well.
\providecommand{\OWNER}[1]{\textcolor{violet}{[\textbf{Owner to pick:} #1]}}
\section{System Design}
\label{sec:design}
\begin{table*}[t]
\caption{What correct training means for each part of a run, the reference each check compares with and its rule. References are of five kinds: the official release's files, the framework's declarations and published recipe, closed-form results, an independent implementation or recomputation, and the run's saved state. None is produced by the code under test.}
\label{tab:truths}
\vskip 0.1in
\centering
\small
\begin{tabularx}{\textwidth}{@{}>{\raggedright\arraybackslash}p{0.14\textwidth}>{\raggedright\arraybackslash}X>{\raggedright\arraybackslash}p{0.2\textwidth}>{\raggedright\arraybackslash}p{0.22\textwidth}@{}}
\toprule
Check & Correct training means & Reference & Rule \\
\midrule
\multicolumn{4}{@{}l}{\emph{Lifecycle (\S\ref{sec:lifecycle})}} \\
Image & the job runs the framework commit under test with the dependency versions it declares & framework declarations & exact \\
\specialrule{0.2pt}{2pt}{2pt}
Collective & each collective computes its defined result on every rank & closed form & exact on every rank \\
\specialrule{0.2pt}{2pt}{2pt}
Data & training sees each conversation as inference does, and the loss covers the text the model generates & official template, tokenizer & exact on every row \\
\specialrule{0.2pt}{2pt}{2pt}
Build and load & the model built and loaded is the released model & official configuration, original checkpoint & exact, and unsupported settings are refused \\
\specialrule{0.2pt}{2pt}{2pt}
Save and resume & a resumed run continues from its saved state & state at the save & bitwise on every rank \\
\specialrule{0.2pt}{2pt}{2pt}
Export and serving & the export is the trained model in the release's format and serves as the original checkpoint does & original checkpoint on the same engine & bitwise at learning rate 0; equal server behavior \\
\specialrule{0.2pt}{2pt}{2pt}
Optimizer step & the loss is the mean cross-entropy over the supervised tokens, and each update is the one the recipe's optimizer settings define & fp32/fp64 recomputation & exact or within rounding bounds; injected faults caught \\
\midrule
\multicolumn{4}{@{}l}{\emph{Architecture (\S\ref{sec:sweep})}} \\
Step & under each supported setting, one step's gradients and loss equal the reference's within noise & model's own implementation (e.g., Hugging Face code) & relative error $\le 4\times$ the larger noise floor, per layer group \\
\midrule
\multicolumn{4}{@{}l}{\emph{Convergence (\S\ref{sec:e2e})}} \\
Convergence & on each representative workload, a run follows its recipe and trains stably to the end & published recipe & \appAconv \\
\bottomrule
\end{tabularx}
\end{table*}

\subsection{Overview}
\label{sec:overview}

Plumbline validates a run at the configuration its user trains with. A fault in any part of a run can make it train or serve the wrong model. Some faults also need a particular setting or many steps to appear. Plumbline therefore checks each part of a run against a trusted reference, under the settings and over the run lengths in which faults appear. Two principles shape the design. The first is correctness first: a system that validates training needs to be trustworthy itself. Each check therefore tests the training a user would run against a reference we can trust, under rules fixed within each system version, a frozen release of Plumbline, so that two results are judged the same way. Each bound sits above measured noise, so noise alone does not fail a check. The second is modularity: each check tests one part on its own. The checks can therefore run one at a time or together as a full scan of a new model, and a failed check already narrows where to look. By design, adding a new framework or model changes how the checks connect to it, not what they test or the rules they apply.

The checks sit at three levels (Fig.~\ref{fig:system}), and Table~\ref{tab:truths} lists the reference and the rule of each. Lifecycle checks cover the parts that surround the forward and backward pass and determine what the model is trained on and what is served (\S\ref{sec:lifecycle}). Architecture checks cover the forward and backward pass under the supported training settings, because new layer types can take different code paths from one setting to another (\S\ref{sec:sweep}). Convergence checks cover whole runs, for faults that need the full model or many steps (\S\ref{sec:e2e}). When a check or a run fails, a diagnosis searches for its root cause with the same checks (\S\ref{sec:diagnosis}). Each check passes or fails by the rules of \S\ref{sec:criteria}. Because the checks are independent and their rules fixed, a coding agent can compose them for each use, while a human reviewer approves its plans and confirms each problem (\S\ref{sec:agents}).

\subsection{Modular Checks and Diagnosis Tools}
\label{sec:components}
Each check compares as exactly as correct training allows. Most lifecycle stages have one correct output and are matched exactly, one training step passes within a bound above measured noise, and a whole run, with no reference run yet, is checked for following its recipe and training stably to the end.

\subsubsection{Lifecycle Correctness}
\label{sec:lifecycle}
\begin{figure*}[t]
\centering
\includegraphics[width=\textwidth]{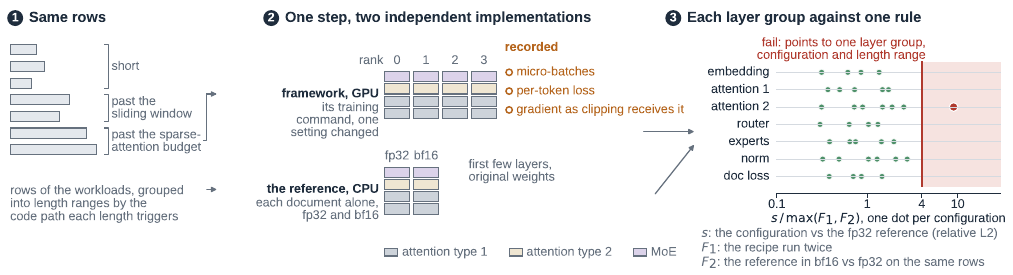}
\caption{The step check. The same rows train one step in the framework's training command with one setting changed and in
the reference implementation (here the model's Hugging Face code) on CPU. A layer group fails when
$s > 4\max(F_1, F_2)$, so each failure points to one layer group, configuration and length range. Dot positions are
illustrative.}
\label{fig:stepcheck}
\end{figure*}
A falling loss curve shows that the model is fitting the supervised tokens it receives. The curve does not reveal whether the data pipeline produces the right tokens and labels, whether checkpointing preserves the weights and training state through loading, resuming and export, or whether each optimizer step normalizes the loss and updates the parameters correctly. These stages run on an environment: the container image and the cluster's communication stack. A fault in the environment or in these stages can silently corrupt what the model learns or serves, or hang the job.

\paragraph{Environment: image versions and collective communication.}
A container image fixes the framework and many separately versioned libraries. A library older than the framework code expects can change results without an error. A faulty network plugin can hang a multi-node run. The image check and the collective check run in the job's own environment: the first in its container image, the second on its nodes and network settings.

\paragraph{Data: chat template, tokenization and loss mask.}
Training learns only from the tokens that the loss mask selects. New chat templates add reasoning blocks and tool turns, which the framework can format, tokenize or mask differently from the official template without an error. The data check runs every row of each representative workload (\S\ref{sec:e2e}) through the framework's data pipeline on CPU. A row passes when its tokens and labels equal those of a reference built from the official chat template and tokenizer.

\paragraph{Checkpoint conversion and loading: build, load, save and resume, export and serving.}
Weights are converted several times between the original checkpoint and the serving engine. A fault in a conversion can load wrong weights, resume from another state or serve a different model, and usually raises no error. The build and load checks compare the framework's model with the official configuration and its weights with the original checkpoint. Every parameter is filled with NaN before the load, so a skipped one shows. The save-and-resume check compares each rank's state after the resumed run's load with its state right after the save, bitwise. The export check first trains one step at learning rate 0, so the exported weights must equal the original checkpoint's. The serving check compares a server on the export with the same engine on the original checkpoint.

\paragraph{Training status: loss normalization and optimizer update.}
Within a step, the loss is normalized by the number of supervised tokens across data-parallel ranks and gradient-accumulation micro-batches, and each parameter group is updated with its own learning-rate and weight-decay settings. An error in either changes what is learned without stopping training. The optimizer-step check recomputes both from the framework's logits and gradients, the loss in fp32 and the update in fp64.

\subsubsection{Architecture Correctness}
\label{sec:sweep}
New layer types, such as linear or sparse attention and new expert routing, can take different kernel paths by training setting and sequence length, so a layer that is right under the recipe may be wrong under another setting a user chooses. The step check (Fig.~\ref{fig:stepcheck}) therefore sweeps configurations. Each is the recipe with one setting changed to another that the framework documents as supported for the model. A few configurations combine settings to test how they interact. Staying within documented support keeps the sweep to configurations a user could run. A setting documented as unsupported is expected to be refused. Each configuration trains one step on fixed rows from the representative workloads, grouped into length ranges, one for each code path that sequence length selects and one for short rows.

To keep each configuration and its reference small, the check uses the first layers of the released model, chosen to include every layer type with its real weights. The reference is the model's own implementation from its release, at a version with its known fixes. For the models we study, this is their Hugging Face code. The reference runs in fp32 on CPU with each document alone, so it shares no kernel, parallel layout or packing with the framework. When the framework runs the same reference code, a bug in that code is outside what this check can detect.

A layer group holds the parameters of one model component, such as attention or the experts. For each group, the check computes
\[
s = \frac{\|g - g_{\mathrm{ref}}\|_2}{\|g_{\mathrm{ref}}\|_2}
\]
over the same gradient elements of the configuration ($g$) and the fp32 reference ($g_{\mathrm{ref}}$). The group fails when ${s > 4\max(F_1, F_2)}$, $4\times$ the larger of two noise floors: the run-to-run floor $F_1$, which is $s$ between two runs of the unchanged recipe, and the precision floor $F_2$, which is $s$ between the reference's bf16 and fp32 runs. The worst document's per-token loss is judged the same way.

\subsubsection{Convergence Correctness}
\label{sec:e2e}
Some faults need the full model or many steps to appear: effects of depth, memory pressure, the number and timing of collectives, and errors that build up between ranks. The convergence checks therefore run the full recipe. The published recipe first runs as it is on a small dataset such as GSM8K~\citep{cobbe2021gsm8k}, so a problem in the recipe itself is seen before any change. A 100-step run then trains on each representative workload: general chat (T\"ulu~3, \citealp{lambert2024tulu3}), thinking-mode reasoning (Dolci-Think-SFT, \citealp{olmo2025olmo3}) and long-document question answering (LongReward-10k, \citealp{zhang2024longreward}). The workloads differ in rendering and sequence length, so faults that depend on the form or length of the data have a workload that exercises them.

A run passes when its records show it followed its recipe and trained stably. These checks catch faults that stop, stall or destabilize a run. The comparison with an accepted earlier run of the recipe is recorded but not yet judged.

\subsubsection{Diagnosis Tools}
\label{sec:diagnosis}
A failed run or check shows where training went wrong but often not why: a fault on one rank appears as a wait on the others. Five tools record what a run did: our run reader, PyTorch's flight recorder, memory snapshot and profiler, and a Python stack dump. Except for the run reader, their debug settings add overhead, so they are turned on only in a separate rerun of the failed job. A diagnosis combines these records with the checks themselves, rerun at smaller scale or with one setting changed to bisect the fault, and with the code at the commit under test (\S\ref{sec:cs1}). Each diagnosis ends with the cheapest check that catches the problem again, written per component type so it carries to the next model.

\subsubsection{Pass/Fail Rules}
\label{sec:criteria}
A value that floating-point rounding does not change matches its reference exactly. Other values pass within a bound set from measured noise or above that rounding. Where a bound could hide a fault, the check also injects that fault into a copy of its input and passes only when it catches it. A check whose comparison cannot be completed reports an inconclusive or not-evaluated result, never a pass.\appAsent

\subsection{Operation with Agents}
\label{sec:agents}
\begin{figure}[t]
\centering
\includegraphics[width=\columnwidth]{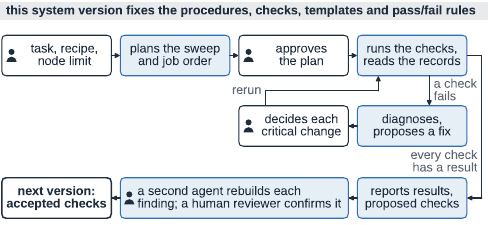}
\caption{How the agent and the human reviewer work within one system version. Tinted steps are the agent's, steps with the person icon the human reviewer's. A failed check goes to diagnosis, and the human reviewer decides each critical change before a rerun. Accepted checks enter the next version.}
\label{fig:agents}
\end{figure}
A full scan of a new model involves many checks, configurations and reruns. The design lets a coding agent operate the scan (Fig.~\ref{fig:agents}). Because each check answers one question with a rule fixed per system version, the agent chooses what to run but not how a result is judged. A diagnosis reruns checks, so the agent can also search for a failure's cause by the same means.
Each system version fixes the procedures, the checks, the per-framework templates that launch and record each check, and the pass/fail rules. Within a version, the agent chooses the configurations and job order, runs the checks, and proposes hypotheses and changes from the records the jobs write. The human reviewer sets the task, approves the agent's plan and decides each critical change, such as a rerun outside the plan or a fix that changes what a check validates. A second agent rebuilds each finding from the raw records, and the human reviewer confirms it.

Diagnoses feed back into the design: a check that a diagnosis names and the human reviewer accepts enters the next system version, which re-evaluates the stored records. The checks thus grow with each problem found, while every result stays tied to the version that judged it. 
\section{Use Cases}
\label{sec:cases}

\subsection{Overview}
\label{sec:cases-overview}
\begin{table}[t]
\caption{The models validated, each on AutoModel, Megatron-SWIFT and Miles except gpt-oss-20b, on Miles only. All five are mixture-of-experts (MoE) models. Model-specific architecture lists the parts beyond a standard MoE transformer block, with multi-token prediction (MTP), and the release's low-precision weight formats. Nodes is the validated training scale: the number of nodes, each with eight B200 GPUs, in the training configurations that onboarding recommends for the model; a range spans the model's workloads and the frameworks' recipes.}
\label{tab:models}
\vskip 0.1in
\centering
\small
\begin{tabularx}{\columnwidth}{@{}>{\raggedright\arraybackslash}p{0.355\columnwidth}>{\raggedright\arraybackslash}X>{\raggedleft\arraybackslash}p{0.1\columnwidth}@{}}
\toprule
Model & Model-specific architecture & Nodes \\
\midrule
gpt-oss-20b~\citep{openai2025gptoss} & MXFP4 experts & 1 \\
\specialrule{0.2pt}{2pt}{2pt}
Qwen3.8-Flash-Next~\citep{qiu2026qwen38next} & linear attention, sparse attention and its indexer, hyper-connections, n-gram table, MTP & 4 \\
\specialrule{0.2pt}{2pt}{2pt}
DeepSeek-V4-Flash~\citep{deepseekai2026v4} & sparse attention, hyper-connections, MTP, FP8/FP4 weights & 8 \\
\specialrule{0.2pt}{2pt}{2pt}
GLM-5.3-Flash~\citep{glm5team2026glm5} & sparse and linear attention, hyper-connections, MTP, FP8 weights & 8--12 \\
\specialrule{0.2pt}{2pt}{2pt}
DeepSeek-V4.1-Flash~\citep{deepseekai2026v41flash} & sparse attention, Engram tables, hyper-connections, DSpark draft layers, FP8/FP4 weights & 8--16 \\
\bottomrule
\end{tabularx}
\end{table}
At Amazon, Plumbline runs in production in two agentic use cases, one for each way modularity lets its checks run (\S\ref{sec:overview}). The two use cases cover LLM training at scale across 3 frameworks, AutoModel~\citep{nemo-automodel}, Megatron-SWIFT~\citep{zhao2025swift} and Miles~\citep{miles-report}, and 5 models (Table~\ref{tab:models}). For a newly released model, a fault can sit in any part of a run, so Day-0 onboarding (\S\ref{sec:cs2}) applies the full scan before teams train it. A failed or implausible result narrows the search, so root-cause tracing (\S\ref{sec:cs1}) uses a few checks and diagnosis tools to find its cause.

From September to October 2026, Day-0 onboarding found 61 problems\appBcite{} at the versions we tested, counting one root cause on one framework as one problem. Root-cause tracing located the cause of each and confirmed a fix or workaround. Teams that use these models can apply the fixes in advance, rather than after a run fails or a trained model underperforms. Of the 61, 42 are fixed or reported upstream, 8 of them first raised by us.

\subsection{Use Case 1: Day-0 Onboarding}
\label{sec:case-day0}\label{sec:cs2}
Teams want to train a newly released model as soon as a framework supports it, with their own data and settings. A new model brings new layer types, weight formats and a chat template that few training runs have yet exercised. Problems can then lie in any part of a run, appear only under some settings or workloads, or be silent: the run goes on with no error and a falling loss. Day-0 onboarding finds such problems, each with a fix or workaround, before a team commits GPUs to its full-scale training.

\begin{table*}[t]
\caption{The 61 problems of \S\ref{sec:cases-overview} by failure pattern and by the part of the full scan that found each first. A cell counts the problems of that pattern that the part found first; silent problems are in parentheses. Code reading: problems the agent saw in the framework's code while building the scan, including one first found by the CPU dry run of a job.}
\label{tab:s4d0-patterns}
\vskip 0.1in
\centering
\small
\begin{tabular}{@{}lrrrrr@{}}
\toprule
Failure pattern & Lifecycle & Architecture & Code reading & Convergence & Total \\
\midrule
Tokens or loss masks differ from the release & 11 (11) & --     & --    & 1      & 12 (11) \\
Layers break under parallelism, packing or other kernels & --      & 9 (5)  & 1 (1) & 4      & 14 (6) \\
Failures depend on sequence length & --      & --     & --    & 9 (1)  & 9 (1) \\
Optimizer updates are lost or misapplied & 3 (3)   & 1 (1)  & --    & 3      & 7 (4) \\
Resumed state differs from the checkpoint & 5 (5)   & --     & --    & --     & 5 (5) \\
Conversion alters the checkpoint & 7 (4)   & --     & 2 (1) & 1      & 10 (5) \\
Multi-node jobs hang or fail to launch & --      & --     & 1     & 3      & 4 (0) \\
\midrule
Total                                 & 26 (23) & 10 (6) & 4 (2) & 21 (1) & 61 (32) \\
\bottomrule
\end{tabular}
\end{table*}
\subsubsection{Workflow}
\label{sec:cs2-workflow}
\begin{figure}[t]
\centering
\includegraphics[width=\columnwidth]{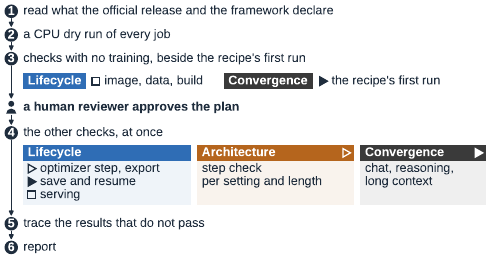}
\caption{One onboarding cycle, from reading what the official release and the framework declare to the report. The numbered steps go from top to bottom, and the jobs within a step run at once. The person icon marks the human reviewer's approval of the plan, which the step after it waits for. Step 5 hands the results that do not pass to root-cause tracing (\S\ref{sec:cs1}). Checks are grouped, colored and marked by the training they need as in Fig.~\ref{fig:system}; the collective and load checks run when their trigger holds, such as a first multi-node run or a quantized release.}
\label{fig:s4d0-cycle}
\end{figure}

Each task names a framework--model combination, its commit and a container image. From what the official release and the framework's documentation state about training, the agent builds the scan: the lifecycle checks, the architecture checks under each supported setting and length range, and the convergence checks on the representative workloads (Table~\ref{tab:truths}). A CPU dry run of each job, the checks that need no training and the recipe's first run start the cycle (Fig.~\ref{fig:s4d0-cycle}). Once these jobs pass and the reviewer approves the plan (\S\ref{sec:agents}), the other checks run at once.

\subsubsection{Results: What the Full Scan Found}
\label{sec:cs2-results}\label{sec:cs2-example}

The 61 problems of \S\ref{sec:cases-overview} fall into seven failure patterns, each named by what goes wrong in training (Table~\ref{tab:s4d0-patterns}). Each part of the scan found some problems first, and in six of the patterns one part found most of them. The lifecycle and architecture checks, which compare each part of training with its reference, found 29 of the 32 silent problems first. At least 16 of the 32 were in runs that passed their convergence checks, so these runs did not show them. The convergence checks found 21 problems first, only one of them silent, because they catch faults that stop, stall or destabilize a run.

\paragraph{Tokens or loss masks differ from the release.} These problems arise where the data pipeline renders, masks or tokenizes a new model's conversations differently from the official chat template and tokenizer. The difference can lie in any detail, such as a turn marker or a reasoning block. Each changes the tokens or the supervised positions. In the silent ones, the rest of each conversation is still trained, so the loss falls as usual. For example, the loss mask of GLM-5.3-Flash left out the token that ends a reply, so the model was not trained to stop. The data check found most of these problems first, before any training run.

\paragraph{Layers break under parallelism, packing or other kernels.} A layer that matches its reference implementation can still fail when its computation is partitioned across ranks, when documents are packed into one sequence, or when another kernel runs. If the layer combines partial results across ranks wrongly or lets a document read across its boundary, only some parameters or documents are affected, so the loss still falls. If the setting selects a kernel that lacks support for the layer, the run stops. For example, the query-normalization weights of DeepSeek-V4-Flash received $8\times$ their gradient at tensor parallelism 8, because a gradient already reduced across tensor-parallel ranks was all-reduced again. The step check found most of these problems first.

\paragraph{Failures depend on sequence length.} Parts of training behave differently at particular sequence lengths: long sequences, very short documents, or lengths that differ between batches or ranks. This behavior covers memory use, kernels and their compilation, and rank synchronization. The resulting failures stop the run, slow it, make the loss NaN, or leave parameters without gradients. For example, the sparse-attention indexer of Qwen3.8-Flash-Next scored every token against every block of the sequence in fp32, so its activations grew with the square of the sequence length. On a workload whose longest sequence has 67,877 tokens, the run ran out of memory at step 2. The convergence checks found all of them first.

\paragraph{Optimizer updates are lost or misapplied.} These problems came from two sources: memory-saving options and a new model's auxiliary modules. Keeping weights, gradients or optimizer state in lower precision or on the CPU saves memory but can lose small updates or fail on mixed-precision parameters. A frozen module can still be updated or contribute to the loss. Lost or wrong updates leave the loss falling, while a failure on mixed parameters or a frozen module stops the run at its start. For example, with AdamW updating bf16 weights in place at a learning rate of $5\times10^{-6}$, 92\% of sampled weights that a step should change kept their value, while the loss looked normal. The build, optimizer-step and step checks found the silent ones first.

\paragraph{Resumed state differs from the checkpoint.} A resume loads each component of the training state separately, so one component can differ from how it was saved. Here, these were progress counters, such as the learning-rate schedule, the data position and the optimizer step, and a sharded embedding table. None of these problems raises an error, and the resumed run keeps training. For example, with its state offloaded to the CPU, the optimizer of GLM-5.3-Flash did not restore its step count after a resume. The resumed run's loss then differed from an uninterrupted run's by up to 3.7\%, while repeated uninterrupted runs differed by at most 0.02\%. The save-and-resume check found all of them first.

\paragraph{Conversion alters the checkpoint.} The model is converted between checkpoint formats twice: when the original checkpoint is loaded for training, and when the trained model is exported for serving. Whatever a conversion does not map, such as a new weight format, tensor, config field or file, is dropped, misread or misnamed. The silent ones do not show during training: an export is served only after training, and a model loaded with wrong weights still trains with a falling loss. For example, loading the MXFP4 checkpoint of gpt-oss-20b transposed the down projections of all 32 experts. The matrices are square, so no shape check caught the transposition. A run from this load started at a loss of 17.08 instead of the 1.48 of a correct load. The run's loss still fell, and the run passed the convergence checks, which judge whether a run trains stably, not its loss value. The export check found most of these problems first. Reading the conversion code found the transposed experts, and the step check confirmed them.

\paragraph{Multi-node jobs hang or fail to launch.} The remaining problems lie outside the model's layers, in cross-node communication, compilation on every rank and the multi-node launch. Such failures show as a hang that logs no error until a timeout, often with the GPUs fully utilized, or as a launch that stops before training. For example, the first multi-node run of Qwen3.8-Flash-Next never finished step 0: every GPU was busy, no line was logged, and one all-to-all of the backward pass stayed pending on a single rank. The convergence checks found the hangs first, and the CPU dry run found a failed launch before any GPU job.

\subsection{Use Case 2: Root-Cause Tracing}
\label{sec:case-rootcause-r4}\label{sec:cs1}
Root-cause tracing starts whenever a check does not pass, a training run fails, or a result is implausible. Each such failure narrows where to look but seldom shows the cause. Testing each hypothesis by rerunning the job at its scale is slow and costly (\S\ref{sec:motivation}). Tracing therefore starts from what the failed job recorded and tests its hypotheses with the diagnosis tools and a few checks at smaller scale.

\subsubsection{Workflow}
\label{sec:cs1-workflow}
\begin{figure}[t]
\centering
\includegraphics[width=\columnwidth]{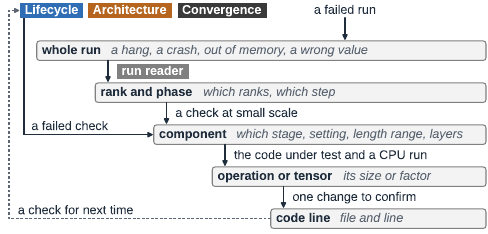}
\caption{How a diagnosis narrows a failure, level by level: whole run, rank and phase, component, operation or tensor, code line. Beside each step down is the tool or test that takes it: the run reader, a check at small scale, the code under test and a CPU run, and one change to confirm. A failed run enters at the top, a failed check at the component it covers. Dashed: the check named for next time feeds back into the three levels.}
\label{fig:cs1-loop}
\end{figure}

From the records that the failed job wrote on every rank, the agent forms a hypothesis and narrows it level by level, from the whole run to a code line (Fig.~\ref{fig:cs1-loop}). At each level the agent chooses the cheapest test that can reject the hypothesis, such as a check on fewer nodes or a CPU run of one operation. The failed job, rerun with one change and no debug tools, then confirms the cause or the fix. By design, each diagnosis ends with a check that catches the problem next time (\S\ref{sec:diagnosis}). When no hypothesis is left that one job can test, the human reviewer decides the next step.

\subsubsection{Results: From Symptom to Root Cause}
\label{sec:cs1-results}
Of the 61 diagnoses, 50 reached the code at the commit under test, the lowest level in Fig.~\ref{fig:cs1-loop}. The descents used the failed job's records, the diagnosis tools and a few tests at smaller scale or on CPU. The cause often lay far from the symptom: a hang from a forward-only kernel selected for short documents, and an illegal memory access from bf16 and fp32 gradients mixed in one fused kernel.

\paragraph{Case study: an out-of-memory error.}\label{sec:cs1-oom}
In one diagnosis, a single-node check and a CPU run explained an eight-node out-of-memory error. The recipe's first run of DeepSeek-V4-Flash trained on rows of a few hundred tokens, but on a representative workload, with rows of up to 4,914 tokens, the run stopped in its first backward pass.

\begin{figure}[t]
\centering
\includegraphics[width=\columnwidth]{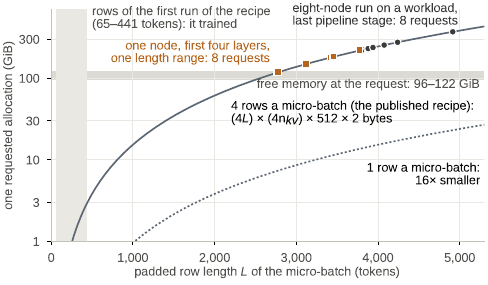}
\caption{The 16 logged out-of-memory requests of the eight-node run's last stage (circles) and of the step check's length range for long rows, on one node (squares). Each is placed at the integer row length $L$ at which the gather's backward, $(b \cdot L) \times (b \cdot n_{\mathrm{kv}}) \times 512$ bf16 rounded up to 2~MiB, equals it exactly. On one node, each such $L$ is the token count of a row in the batch. On eight nodes, the largest is the workload's longest row, 4,914 tokens. Lines: 4 rows a micro-batch (solid) and 1 row (dotted).}
\label{fig:cs1-p50}
\end{figure}

The run reader showed every rank of the last pipeline stage had requested one allocation of hundreds of GiB, more than twice its free memory, with no log line naming the operation. An allocation that large, growing with the rows, pointed to an operation whose memory is superlinear in row length.

The step check's length range for long rows reproduced the error on a single node in about 3 minutes, also under other parallel and recomputation settings. The failure therefore depends on row length, not on pipeline parallelism or eight-node communication. Reading the code at the commit under test located the operation. The sparse attention gathers each query's keys and values from an expanded view. For micro-batches of $b$ rows of $L$ tokens, each row with $n_{\mathrm{kv}}$ key-value entries of 512 values, the gather's backward allocates the whole view's gradient of $(b \cdot L) \times (b \cdot n_{\mathrm{kv}}) \times 512$ bf16 values. A small CPU run confirmed this dense gradient, and with the allocator's rounding to 2~MiB the formula matched all 16 logged requests exactly (Fig.~\ref{fig:cs1-p50}). Locating the cause took one check of about half a node-hour.

\subsection{Across Use Cases}
\label{sec:across}
Both use cases rely on the checks' verdicts, and together they show where problems sit.

\paragraph{Calibration of the checks.}
The lifecycle checks compare exactly or bitwise. The optimizer-step check compares the loss and each update with a high-precision recomputation, within rounding tolerances that an injected fault tests on every run (\S\ref{sec:criteria}). The step check instead passes within a multiple of measured noise. We examined the step checks of 3 models on 3 frameworks. Where a run had no known defect, the unchanged recipe's error stayed at or below $3.44\times$ the larger noise floor $\max(F_1, F_2)$, under the bound of 4. Where a step carried a known defect and failed, the error was at least $4.35\times$ that floor.

\paragraph{Problems across models and frameworks.}
Of the 61 problems, 28 appeared in a model-specific part that Table~\ref{tab:models} lists, such as its attention variant or low-precision weights. Counted by where the cause lies, 27 lay in the framework's dependencies rather than its own code, most in Megatron-LM or in the bridges that implement new models and checkpoint conversion for it. These dependencies are released apart from the framework, so its pinned versions can lack support for a new model or an upstream fix. Where a cause lies decides which combinations a problem reaches. A cause in the model can recur on each framework that trains it, as when two frameworks' exports left out the n-gram table of Qwen3.8-Flash-Next, and a cause in a framework's shared code can recur on each of its models, as when one export wrote no generation configuration for three of them. 
\section{Limitations}
\label{sec:limitations}
\paragraph{Each check has a defined scope.} A check gives a verdict where it has a trusted reference (\S\ref{sec:overview}). DeepSeek-V4.1-Flash, for example, has no Hugging Face implementation at the versions we tested, so the step check leaves its layers unjudged until one is released. We consider two fallbacks, each with a narrower scope: its inference code, which covers only the forward pass, and a comparison of each setting with the framework's unchanged recipe, which shares the code under test. Plumbline records the latter for reference, without a verdict. The step check runs one step of a model's first layers, so faults that need the full model or many steps are left to the lifecycle and convergence checks.

\paragraph{A human reviewer confirms each problem.} Today the agent operates Plumbline, and a human reviewer approves its plans and confirms each problem before it reaches a team (\S\ref{sec:agents}). We are moving toward full automation. Each system version adds the checks and procedures that the reviewer accepted, so fewer decisions are left to the reviewer, and onboarding more models, frameworks and use cases makes the checks and their rules more general. Our next step is for the agent to propose and test new checks itself.

\section{Related Work}
\label{sec:related}

\paragraph{Training-correctness checking.} A falling loss does not show whether a training run computes what the
released model defines.
The training debuggers SageMaker Debugger~\citep{rauschmayr2021smdebugger}, AUTOTRAINER~\citep{zhang2021autotrainer}
and DeepDiagnosis~\citep{wardat2022deepdiagnosis} watch a run's tensors and metrics for symptoms such as vanishing
or exploding gradients.
The closest systems compare training with a counterpart. TrainCheck~\citep{jiang2025traincheck}
checks a running job against invariants inferred from high-quality example pipelines.
TrainVerify~\citep{lu2025trainverify} proves a parallel execution plan equivalent to the single-device logical plan.
Entangle~\citep{wang2026entangle} verifies that a distributed model refines its sequential model.
nnTrace~\citep{jiang2026nntrace} aligns the tensors of one distributed iteration with a trusted reference run.
OpGuard~\citep{zhou2026opguard} localizes a production incident to the first operator that differs bit for bit from a
comparable run.
These counterparts come from training code: invariants of example programs, or another plan, setting or run of a
framework.
When a framework first supports a new model, its settings and runs share the new model code, so comparing them can
miss a fault in that code. Our references for that
code come from the official release.

\paragraph{Framework tests, checkpointing and differential testing.}
Megatron-LM~\citep{mlm-tests} tests short and resumed runs of its own model configurations against reference runs,
and Megatron-Bridge's verification card~\citep{mbridge-card} requires a one-prompt logit comparison with the Hugging
Face model and a resumed run within 1\% of the uninterrupted run.
ByteCheckpoint~\citep{wan2025bytecheckpoint} and Universal Checkpointing~\citep{lian2025ucp} reshard checkpoints
across parallel layouts and compare the loss of a resumed run with that of the original run.
CRADLE~\citep{pham2019cradle}, Muffin~\citep{gu2022muffin}, FreeFuzz~\citep{wei2022freefuzz} and
TitanFuzz~\citep{deng2023titanfuzz} compare models or API calls across library backends or devices,
NNSmith~\citep{liu2023nnsmith} generates valid models to test compilers, and D$^3$~\citep{wang2025d3} requires a
generated model trained under different distributed settings to match its non-distributed run. CPU-GPU comparison
missed 30 of the 31 silent bugs that TransFuzz found~\citep{zhang2026transfuzz}.
Plumbline instead runs each framework's training code on the released model: the step check compares one step's
gradients with the reference implementation's under each supported setting, and the save-and-resume check compares
each rank's saved and restored states bitwise.

\paragraph{Training failures at scale and their diagnosis.} MegaScale~\citep{jiang2024megascale} and
ByteRobust~\citep{wan2025byterobust} detect, localize and recover from failures in large jobs.
Mycroft~\citep{deng2025mycroft} traces collective communication to explain stalls and slowdowns. These systems isolate
the component or code update at fault during a long job.
When a run fails, our diagnoses follow similar per-rank records into the code at the commit under test.
SuperBench~\citep{xiong2024superbench} validates nodes before jobs use them, and Aegis~\citep{dong2025aegis} and
GUARD~\citep{liu2026guard} add such checks to their runtime monitoring.
Our checks test each part of the training software before such a job.

\paragraph{Validation and reproducibility.} MLPerf Training~\citep{mattson2020mlperf} requires closed-division
submissions to be mathematically equivalent to a reference implementation. Even with fixed random seeds, identical
training runs differed in accuracy by up to 2.9 points~\citep{pham2020variance}, and FLEX~\citep{dutta2021flex}
derives the bounds of ML tests from the tail of repeated runs. Our step check also sets its tolerance from measured
noise. The ML Test Score~\citep{breck2017mltestscore}, schema-based data validation~\citep{polyzotis2019datavalidation},
model assertions~\citep{kang2020assertions} and ease.ml/ci~\citep{renggli2019easeml} test a team's own ML pipeline
against stated conditions. Plumbline states fixed conditions for the frameworks that teams build on.
In concurrent work, Cross-Stack Validation~\citep{tran2026crossstack} uses an independent training stack as the oracle
for LoRA training of Qwen3-0.6B, and its fault with the largest effect on the trained model lay in how the training
data was formatted, which our data check compares with the official chat template and tokenizer.

\paragraph{LLM agents for systems debugging and testing.} SWE-bench~\citep{jimenez2024swebench},
KernelBench~\citep{ouyang2025kernelbench} and AIOpsLab~\citep{chen2025aiopslab} judge the patches, kernels and
failure diagnoses that LLMs and agents produce against tests, reference code and injected faults fixed by the
benchmark.
Ekka's agent~\citep{gu2026ekka} maps a serving engine's modules to a reference implementation
and aligns their activations to diagnose silent errors.
TritorX's agent~\citep{hammond2026tritorx} writes Triton kernels for PyTorch ATen operators on Meta's MTIA, each counted as covered when its OpInfo tests pass.
In concurrent work, ForgeTrain's coding agent~\citep{he2026forgetrain} builds a training engine that must match a trusted framework bit for
bit on anchor tensors before optimizing it, with no human in the loop.
OpGuard Harness~\citep{opguard-harness}, a code release, lets a coding agent
investigate two diverging runs and reports no pass/fail verdict.
Plumbline's agent chooses which checks to run under a human reviewer's approval, and the pass/fail rules come from the system version, not the agent.
\section{Conclusion}
\label{sec:conclusion}
A falling loss does not show that an LLM training run is correct. Of the 61 problems that Plumbline found, 32 were silent, and checks that compare one part of training with a trusted reference found 29 of these 32 first. Many problems lay in what is new in each model or in the dependencies a framework builds on, and a problem found on one combination points to others worth checking. Plumbline puts this into practice: its checks cover a run at three levels, its diagnosis reuses them to trace a failure to its root cause, and coding agents operate both in production at Amazon. We will release Plumbline as open source.
\clearpage
\bibliography{refs}
\bibliographystyle{mlsys2025}

\section*{AI Usage Disclosure}
Coding agents are part of Plumbline itself
(\S\ref{sec:agents}, \S\ref{sec:cases}). They adapt the checks' launch and record templates to new frameworks, run jobs, choose the settings to test and
trace failed runs. A human reviewer approves each plan and each critical change and confirms every problem we report.
Beyond the system, we used Claude Code~\citep{anthropic2026claudecode} to assist with ideation, literature search, experiment design, parts of the code
(validation code, analysis scripts and figure scripts) and drafts of text, tables and figures. The authors set the
research direction and the design, and decided which experiments to run. They wrote parts of the code, reviewed and tested all of it, checked
every number against the run records and every reference against its source, and revised the text. The authors take
full responsibility for the content. The training workloads are public datasets.

\end{document}